\documentclass[a4paper]{article}
\usepackage{a4wide}
\usepackage{amsmath}
\usepackage{amsfonts}
\usepackage{color}
\usepackage{graphics}
\usepackage{harvard}

\newcommand{\eg}[1]{{\it e.g.\/}\ifx#1.\else\expandafter#1\fi}
\newcommand{\Eq}[1]{(\ref{#1})}
\newcommand{\fig}[1]{fig.~\ref{fig:#1}}
\newcommand{\Fig}[1]{Figure~\ref{fig:#1}}
\newcommand\ie[1]{{\it i.e.\/}\ifx#1.\else\expandafter#1\fi}

\renewcommand{\@}{\partial}
  \newcommand{\<}{\langle}              
\renewcommand{\>}{\rangle}              
\newcommand{\avg}[1]{\overline{#1}}	
\newcommand{\Complex}{\mathbb{C}}	
\newcommand{\const}{\mathrm{const}}
\renewcommand{\d}{\mathrm{d}}
\newcommand{\e}{\ensuremath{e}}		

\renewcommand{\i}{\ensuremath{i}}	
\newcommand{\mx}[1]{\mathbf{#1}}	
\renewcommand{\O}[1]{O\left(#1\right)}  
\newcommand{\Real}{\mathbb{R}}
\renewcommand{\Re}{\mathrm{Re}}
\newcommand{\sign}{\mathop{\mathrm{sign}}}

\newcommand{\A}{\mx{A}}                 
\renewcommand{\a}{\mx{a}}               
\newcommand{\D}{\mx{D}}                 
\newcommand{\Disk}{\mathcal{D}}		
\newcommand{\f}{\mx{f}}			
\newcommand{\h}{\mx{h}}                 
\renewcommand{\H}{H}			
\renewcommand{\L}{\mathcal{L}}		
\newcommand{\Lmov}{\tilde\L}		
\newcommand{\Lp}{\Lmov^{+}}		
\renewcommand{\r}{{\vec r}}             
\newcommand{\R}{{\vec R}}               
\newcommand{\Square}{\mathcal{S}}	
\newcommand{\T}{^T}			
\newcommand{\Tr}{\V}			
\renewcommand{\u}{\mx{u}}		
\newcommand{\U}{\mx{U}}			
\newcommand{\Ulab}{\U}			
\newcommand{\Umov}{\tilde\U}		
\newcommand{\vH}{{\vec{\H}}}            
\newcommand{\V}{\mx{V}}                 
\newcommand{\w}{\mx{w}}			
\newcommand{\W}{\mx{W}}			
\newcommand{\RF}{\W}			
\newcommand{\stept}{{\Delta{}t}}        
\newcommand{\stepx}{{\Delta{}x}}	
\newcommand\theorEps{\varepsilon}
\newcommand\fhnEps{\epsilon}

\newcommand{\myfigure}[3]{
\begin{figure*}[tbp]
\centerline{\includegraphics{#1}}
\caption[]{#2}
\label{fig:#3}
\end{figure*}
}

\begin{document}

\title{Localization of response functions of spiral waves in the FitzHugh-Nagumo system}

\author{
  I. V. Biktasheva$^{1,*}$, 
  A. V. Holden$^2$,
  V. N. Biktashev$^3$
}

\date{\today}
\maketitle

$^1$ Department of Computer Science, University of
Liverpool, Liverpool L69 7ZL, UK

$^2$ School of Biological Sciences, University of Leeds,
Leeds LS2 9JT

$^3$ Department of Mathematical Sciences, University of
Liverpool, Liverpool L69 7ZL, UK

$^*$ Author to whom correspondence should be addressed.

\begin{abstract}
Dynamics of spiral waves in perturbed, \eg\ slightly inhomogeneous or subject to a small periodic external force, two-dimensional autowave media can be described asymptotically in terms of Aristotelean dynamics, so that the velocities of the spiral wave drift in space and time are proportional to the forces caused by the perturbation. The forces are defined as a convolution of the perturbation with the spiral's Response Functions, which are eigenfunctions of the adjoint linearised problem. In this paper we find numerically the Response Functions of a spiral wave solution in the classic excitable FitzHugh-Nagumo model, and show that they are effectively localised in the vicinity of the spiral core.  
\end{abstract}

\section{Introduction}

Autowaves are nonlinear waves observed in spatially distributed media of physical, chemical, and biological nature, where wave propagation is supported by a source of energy stored in the medium.
In a two-dimensional autowave medium there may exist autowave vortices appearing as rotating spiral waves and thus acting as sources of periodic waves. 
Their existence is not due to singularities in the medium but is determined only by development from initial conditions.
In a slightly perturbed medium, \eg\ spatially inhomogeneous or subject to time-dependent external forcing, a spiral wave drifts, \ie\ its core location and frequency change with time \cite{ENS,PRE2000,JOBP,Fast-Pertsov-1990,Pertsov-Ermakova-1988}.

While the hypothesis of re-entry of excitation underlying cardiac arrhythmias belongs to the beginning of the twentieth century, \eg\ \cite{Mines-1913}, the first direct experimental observation of spiral waves was reported in 1960s in a chemical oscillatory medium, the Belousov-Zhabotinsky (BZ) reaction \cite{Zhabotinsky-Zaikin-1971}. That triggered a huge amount of interest and activity in the area. Soon after that spiral waves were observed in a rabbit ventricular tissue \cite{Allessie-etal-1973}, and later in a variety of other spatially distributed active systems: in chick retina 
\cite{Gorelova-Bures-1983}, colonies of social amoebae \cite{Alcantara-Monk-1974}, cytoplasm of
single o\"{o}cytes \cite{Lechleiter-etal-1991}, 
in the reaction of catalytic oxidation of carbon oxide \cite{Jakubith-etal-1990}, rusting of the steel surface in acid with the air 
\cite{Agladze-Steinbock-2000}, 
in liquid crystal \cite{Frisch-etal-1994} and laser \cite{Yu-etal-1999} systems.
On a larger scale, there are waves of infectious diseases travelling
through biological populations
\cite{Carey-etal-1978,Murray-etal-1986}, and spiral galaxies
\cite{Madore-Freedman-1987,Schulman-Seiden-1986}. Yet for experimental
studies of spiral waves dynamics the BZ reaction medium remains the
most favourite.

A common feature of all these phenomena is that they can be mathematically approximated by ``reaction-diffusion'' partial differential equations,
\begin{equation}
\@_t\u = \f(\u) + \D \nabla^2 \u, \quad 
\u,\f\in\Real^\ell,\; 
\D\in\Real^{\ell\times\ell},\;
\ell\ge2,					\label{RDS_org}
\end{equation}
where $\u(\r,t)$ is a column-vector of the reagent concentrations,
$\f(\u)$  of the reaction rates, 
$\D$ is the matrix of diffusion coefficients,
and  
$\r\in\Real^2$ is the vector of coordinates on the plane. 
Since these equations are essentially nonlinear, their spiral wave solutions in general case are studied numerically. 
Thus, given the complexity of the problem, the current understanding of spiral waves is mostly empirical and gives neither possibility for systematic quantitative predictions of the drift, nor general understanding on how to control the smooth dynamics of autowave vortices, which is important for many practical applications. Effective control of re-entry in excitable cardiac tissue will provide a solution to dangerous arrhythmias and fatal fibrillation. 

As a model self-organizing structure, spiral wave demonstrates a remarkable stability, just changing its rotational frequency and core location, \ie\ drifting, in response to small perturbations of the medium. As experiments with BZ reaction medium \cite{Agladze-private} and computer simulations showed spiral waves insensitivity to distant events, it was conjectured \cite{vnbdisser} that the RFs must decay quickly with distance from the spiral wave core, \ie\ spiral waves \emph{look like} essentially non-localized regimes but \emph{behave} as effectively localized particles \cite{swd}. The asymptotical theory of the spiral wave drift, proposed in \cite{Keener-1988,Biktashev-Holden-1995} and shortly described below, is based on the idea of summation of elementary responses of the spiral wave core position and rotation phase to elementary perturbations of different modalities and at different times and places. This is mathematically expressed in terms of the spiral wave \textit{response functions} (RFs) equal to zero in the region where the spiral wave is insensitive to small perturbations.

So far, the response functions have been explicitly found with good quantitative accuracy only for spiral waves in oscillatory medium described by the Complex Ginzburg--Landau Equation (CGLE) \cite{PRE1998,JNMP,swd}. It was shown that the response functions of vortices in the CGLE medium are essentially nonzero only in the vicinity of the core for all sets of model parameters stable spiral wave solution exists for \cite{JNMP,swd}, which explains the localised sensitivity of spiral waves to small perturbations. Most important is the RFs ability to make quantitative prediction of spiral wave drift velocity due to small perturbations of any nature.

Thus, a spiral wave organise the medium dividing it into two unequal parts, the core, events in which are translated throughout the medium, and the periphery, obeying the signals from the core. It creates a \emph{macroscopic} wave-particle dualism as an emergent property of the nonlinear field, when the regime \emph{appears} as a non-localized object filling up all available space, but \emph{behaves} as a localized object, only sensitive to perturbations affecting its core.

Another class of media supporting spiral waves are excitable media. These are even of more interest than the oscillatory ones, due to their role in the cardiac, smooth muscle research and neuroscience.
In order to check localisation properties of the response functions of vortices in excitable media, the RFs need to be found explicitly for a particular excitable model. Hamm \cite{Hamm-1997} tried to find the response functions for the Barkley model
of an excitable system. The obtained response functions were
effectively localised in the vicinity of the spiral wave core, but the accuracy of the
solution was not sufficient to allow it to be used for prediction of the velocity of the spiral wave drift.

In this paper we find the response functions for a spiral wave
solution in the FitzHugh-Nagumo (FHN) model
\cite{FitzHugh-1961,Nagumo-etal-1962} and show that the RFs are
effectively localized in the vicinity of the spiral wave core. The
model parameters were selected to produce an excitable medium with a
rigidly rotating spiral wave. The method of computation is based on
the idea of a moving frame of reference, whose movement is controlled
by the spiral wave solution found in that frame
\cite{Biktashev-etal-1996}.

The FitzHugh-Nagumo model is historically the first simplified model of biological excitation. It has been studied and used as a classic model for computer simulation of spiral wave dynamics for decades, for it captures the key phenomena of the excitable media while consisting of just two partial differential equations, which makes the FHN model easy to study both numerically and analytically,
\begin{eqnarray}
\@_tu_1 &=& \fhnEps^{-1} \left( u_1 - u_1^3/3 - u_2\right) + D_1 \nabla^2u_1 \nonumber\\
\@_tu_2 &=& \fhnEps \left(u_1+\beta-\gamma u_2\right) 
							\label{FHN}
\end{eqnarray}
where $\beta$, $\gamma$, $\fhnEps$ and $D_1$ are parameters.

In the FHN model the variable $u_1$ is the fast variable, corresponding to the voltage in biophysically realistic models of membrane action potential, and $u_2$ does not have any specific physiological interpretation, just plays the role of the slow ``recovery'' variable. The cubic nonlinearity of the system results in the simple N-shape nullcline on the phase portrait and explains the key aspects of excitability. 
Existence of spiral wave solutions in the FHN 
model, their characteristics and behaviour depending on the model parameters have been extensively studied by many 
authors. The classic review on the subject is the Winfree article \cite{Winfree-1991}.

\section{Asymptotic theory of spiral waves dynamics} \label{asymtheory}

\subsection{Initial definitions}

Consider a slightly perturbed ``reaction-diffusion'' system \Eq{RDS_org} in two spatial dimensions, 

\begin{equation}
\@_t\u = \f(\u) + \D \nabla^2 \u + \theorEps \h, \quad 
\u,\f,\h\in\Real^\ell,\; 
\D\in\Real^{\ell\times\ell},\;
\ell\ge2,								\label{RDS}
\end{equation}
where 
$\theorEps \h(\u,\r,t)$ is a small perturbation, and $\r\in\Real^2$.

We assume that unperturbed system \Eq{RDS_org} has solutions in the form of steadily rotating spiral waves, 
\begin{equation} 
\u = \Ulab(\r,t) = \Umov(\rho(\r),\vartheta(\r)+\omega t) .  
                                     \label{SW} 
\end{equation} 
Here 
$$ 
\theta = \vartheta(\r) + \omega t 
$$ 
is a polar angle in the ``corotating'' frame of reference, which is
rigidly rotating with the angular velocity $\omega$, while $\rho(\r)$
and $\vartheta(\r)$ are the polar coordinates in the original
(laboratory) frame of reference.

The unperturbed reaction-diffusion system \Eq{RDS_org}, or \Eq{RDS} with
$\theorEps \h=0$, has an obvious but important symmetry: it is
invariant with respect to the Euclidean group of motions of the plane
$\{\r\}$. Since solution \Eq{SW} at any fixed $t$ is not invariant
against this group, the group ``multiplies'' this solution. That is,
\begin{equation}
    \Ulab'(\r,t) =\Umov(\rho (\r-\R),\vartheta (\r-\R) + \Theta) ,
\end{equation}
where $\Theta = \omega t - \Phi$, is another solution for any constant
displacement vector $\R=(X,Y)^{\dag}$ and initial rotation phase $\Phi$.

Thus, if the unperturbed system has one spiral wave solution, then it
has a whole three-dimensional {\em manifold\/} of such solutions, that are
relatively stable with respect to the shift along the manifold.

\subsection{Finite-dimensional analogy}

The asymptotic theory of drift of spiral waves 
\cite{Biktashev-Holden-1995} was proposed based on the analogy with 
finite-dimension problem of perturbation of an invariant manifold 
(see \fig{manifold}). If a vector field $\f(\u)$ in an 
$n$-dimensional phase space has an invariant
$m$-dimensional manifold $\U(\a)$, $m<n$, stable as a whole, then small
perturbation of this vector field will, under certain conditions,
preserve the invariant manifold, just slightly displacing it,
$\U\mapsto{}\U'$.  Another effect of the perturbation is that the vector
field on the shifted manifold $\A'(\a)$ will be slightly different from
the original one, $\A(\a)$. In practice, the existence of the original
invariant manifold $\U(\a)$ could be due to a symmetry group. In that
case, the flow on that manifold could be in some sense degenerate, and
then the perturbation will remove this degeneracy.
 
\myfigure{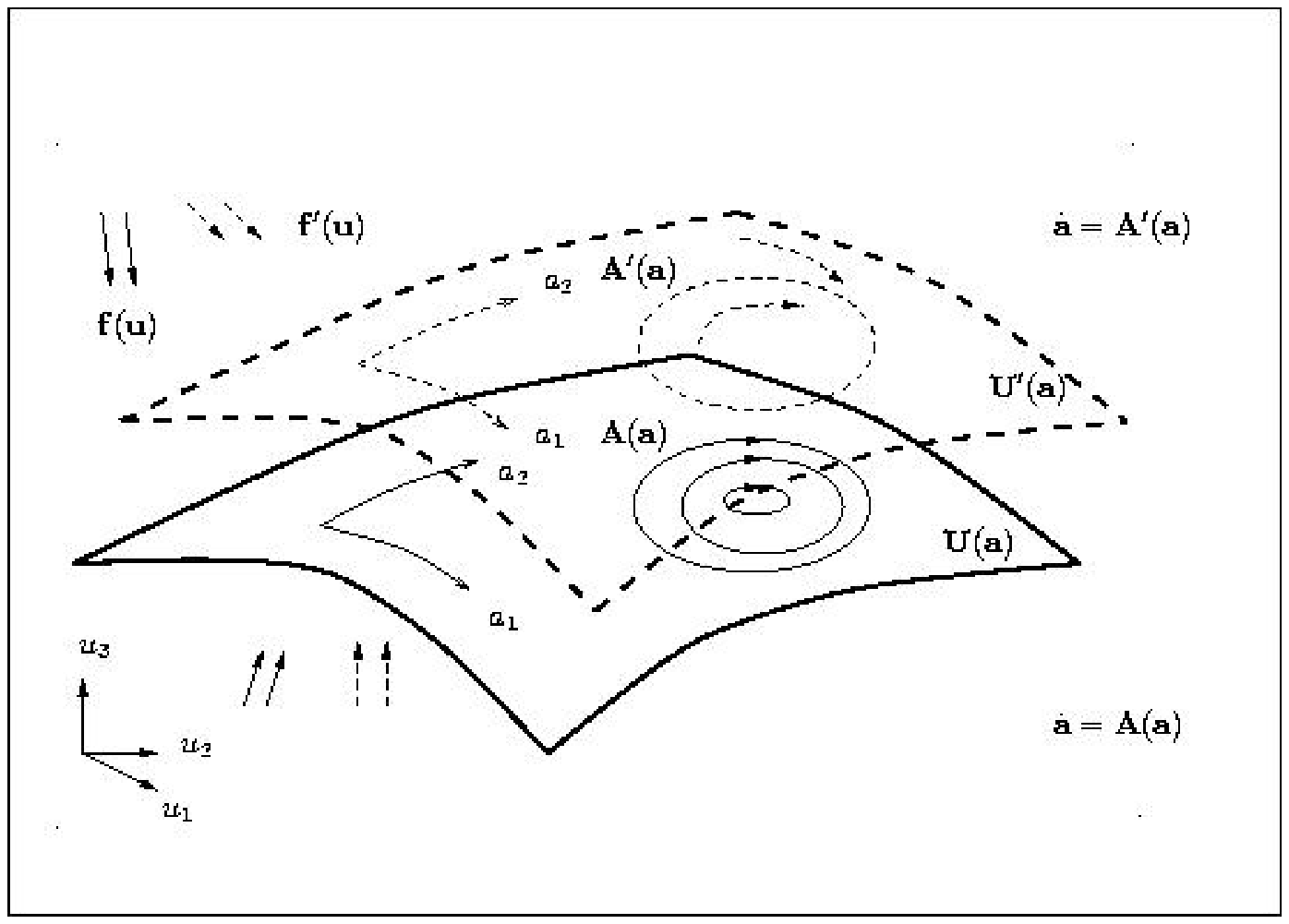}{
  Perturbation of an invariant manifold. Vector field $\f(\u)$ in phase
  space with coordinates $\u$ has an invariant manifold $\U$ with
  coordinates $\a$, and vector field $\A$ on the manifold. Perturbed
  vector field $\f'(\u)$ has a slightly different invariant manifold,
  $\U'$, and a slightly different vector field $\A'$ on it. Original
  objects are shown by solid lines, and perturbed objects by dashed
  lines.
}{manifold}

To compare the two vector fields, on the original manifold and on the
perturbed, we need to relate their coordinate systems $\{\a\}$. A
natural way is to require that the vector connecting two corresponding
points $\U(\a)$ and $\U'(\a)$, would not have a
component along the manifold, \ie\ along any of the tangent vectors
$\V_j(\a)=\@\U/\@a_j$. In other words, it should be orthogonal,
\begin{equation}
\<\W_j(a),\U'(a)-\U(a)\>=0, \quad j=1\dots{}m,
\end{equation}
to the projectors $\W_j(a)$ onto the tangent vectors 
$\V_j(a)$: 
\begin{equation} 
\<\W_j(a),\V_k(a)\>=\delta_{j,k} .  
                    \label{norm}
\end{equation}
These projectors are eigenvectors
of the adjoint linearized matrix $(\@\f/\@\u)\T(\a)$.  The two
effects of the perturbation are produced by its two components, along
and across the manifold, as determined by the projectors $\W_j$.

Thus, if the manifold comprises only non-moving points, 
the tangent component will determine the slow drift along the manifold.  

If this finite-dimensional scheme can be applied to 
spiral waves, the role of the vector field is played by
the reaction-diffusion system, so the phase space is a functional
space. The invariant manifold is the three-dimensional manifold of spiral
waves and is due to a symmetry group, the Euclidean group of the
plane. The coordinates on the manifold are $\R\in\Real^2$, the centre of rotation
of the spiral wave, and $\Theta$, its rotation angle.  The flow on the
manifold is degenerate, as it consists of relatively stable periodic
orbits, which correspond to steady rotation of spiral waves around
fixed centres:
\begin{equation}
  \Theta = \omega t - \Phi, \quad \Phi=\const;\;\R=\const. \label{unptb}
\end{equation}

The perturbation removes this degeneracy, and we observe the 
drift of the spirals. By analogy with the finite-dimensional
case, we expect that the flow on the manifold of spiral waves will be
described by 
\begin{equation}
\@_t\Theta = \omega+\theorEps \H_0(\R,\Theta), \quad
\@_t\R=\theorEps\vH_1(\R,\Theta) ,             \label{ptb}
\end{equation}
where $\H_0$ and $\vH_1$ are ``projections'' of the perturbation
onto the tangent space of the manifold $\U(\a)$. The right-hand sides of \Eq{ptb}
depend on the phase $\Theta$. On the time scale $\theorEps^{-1}$ this
phase oscillates fast; averaging over these oscillations gives motion
equations of the spiral waves, 
\begin{equation}
\@_t\avg{\Theta} = \omega+ \theorEps \avg{\H_0}(\R) + \O{\theorEps^2},\quad
\@_t\avg{\R} = \theorEps \avg{\vH_1}(\R) + \O{\theorEps^2}. 
					\label{Aristotle}
\end{equation}

\subsection{Response functions}

Thus, the finite-dimensional analogy suggests that the dynamics of
spiral waves (perhaps like that of many other dissipative structures) is
described by ``Aristotelean'' mechanics, when the velocity of motion is
proportional to the applied perturbation. The right-hand sides in the
equations, the ``forces'', are projections of the perturbation onto the
corresponding tangent space of the invariant manifold $\U(\a)$. This tangent
space is a linear space, the span of the Goldstone modes, corresponding
to the translations along the symmetry group, at $\R={\bf 0}$ and
$\Theta=0$,
\begin{eqnarray}
\Tr_0 &=& 
  - \omega^{-1} \@_t \Ulab(\r,t) |_{t=0} = 
  - \@_\theta \Umov(\rho(\r),\theta(\r)), \nonumber\\
\Tr_{\pm1} &=& 
  -\frac12 e^{\mp\i\omega t} \left(\@_x\mp\i\@_y\right) \Ulab(\r,t)|_{t=0}  =
  -\frac12 \left(\@_{\tilde{x}}\mp\i\@_{\tilde{y}}\right) \Umov(\r,t) =
  -\frac12 e^{\mp\i\theta} \left(
    \@_\rho\mp\i\rho^{-1}\@_\theta
  \right) \Umov(\rho(\r),\theta(\r)) . \nonumber\\ 
                                        \label{Goldstone} 
\end{eqnarray}
Here mode $\Tr_0$ corresponds to the shift in time (or to what is the
same, rotation in space), and $\Tr_1$ corresponds to the shift in
space.  Tildes in \Eq{Goldstone} designate the corotating frame of
reference, so $\tilde{x}$, $\tilde{y}$ are Cartesian coordinates there
and $\Umov$ is the unperturbed spiral wave solution, which is
stationary in that frame of reference.  We omit the tildes
henceforth for brevity. 

The Goldstone modes are critical eigenfunctions 
\begin{equation}
\Lmov \Tr_n = \i\omega{}n \Tr_n ,\quad n=0,\pm1 
\end{equation}
of the linearized operator $\Lmov$:
\begin{equation}
\Lmov = \D\nabla^2 - \omega\@_\theta 
  + \left.\left({\frac{\@\f}{\@\u}}\right)\right|_{\u=\U(\r)}.
\end{equation}
Here again the tilde at $\Lmov$ reminds that the linearized operator
is considered in the co-rotating frame of reference where it does not
depend on time.  The additive ``$- \omega\@_\theta$'' appears here due
to rotation with respect to the original
system of coordinates.

Thus, for each particular point at the manifold, the projection
operators map the functional space of the perturbations into the
three-dimensional tangent space, and are thus just three functionals.
Since all points of our manifold are equivalent to each other up to a
Euclidean transformation of the plane, it is enough to know the
projection functionals at one point, \ie\ just for one location of the 
spiral wave. This symmetry consideration shows that if the functionals 
$\avg{\H_n}$ are written as integrals, they should have the form:
\begin{equation}
  \avg{\H_n}(t) =
  \e^{ \i n \Phi }
  \oint\limits_{t-\pi/\omega}^{t+\pi/\omega} 
  \frac{\omega\d\tau}{2\pi}   
  \iint\limits_{\Real^2}
  \d^2\r \; 
  \e^{-\i n\omega\tau} 
  \left\<
  \RF_n\left(\rho (\r-\R),\vartheta (\r-\R)+\omega\tau-\Phi\right),\h
  \right\> ,
  					\label{forces}
\end{equation}
where
\begin{eqnarray}
&& \h=\h(\Ulab(\r,\tau),\r,\tau),        \nonumber \\
&& \R=\R(t),                          \nonumber  \\
&& \Phi=\Phi(t),                      \nonumber  \\ 
&& \avg{\H_1}=(\avg{\vH_1})_x + \i(\avg{\vH_1})_y.
\end{eqnarray}
The kernels $\RF_n$ of the integrals (\ref{forces}) are  
eigenfunctions 
\begin{equation} 
\Lp \RF_n = -\i\omega n\RF_n, \quad n=0,\pm1.  
                      \label{RFn} 
\end{equation} 
of the adjoint linearized operator considered in the co-rotating frame
of reference:
\begin{equation} \Lp = \D\nabla^2 + \omega\@_\theta + 
\left.\left({\frac{\@\f}{\@\u}}\right)^{+}\right|_{\u=\U(\r)} .  
                               \label{Lp}
\end{equation}
As in the finite dimensional example, we assume here that $\RF$ 
are normalized in such a way that \Eq{norm} is satisfied. 

The functions $\RF_n$ are called {\em the response functions\/} (RFs) 
of the spiral wave. Folloring the analogy with the Goldstone modes, 
$\RF_0$ defines the shift in time (or the turning in space), and $\RF_1$ 
defines the shift in space. So $\RF_0$ is called the temporal or 
rotation RF, and $\RF_1$ is called the spatial or  
shift RF.

\section{Mathematical formulation of the problem for the FitzHugh-Nagumo model}

The problems \Eq{RDS_org}, \Eq{SW} and \Eq{RFn}--\Eq{Lp} for the FitzHugh-Nagumo system 
\Eq{FHN} take the form
\begin{eqnarray}
&&\fhnEps^{-1} \left( U_1 - U_1^3/3 - U_2\right) + (D_1\nabla^2-\omega \@_\theta)U_1 = 0, \nonumber\\
&&\fhnEps \left(U_1+\beta-\gamma U_2\right)
					-\omega \@_\theta)U_2  = 0,  \label{SWFHN}\\
&&\left( \fhnEps^{-1}(1-U_1^2)+\omega(in+\@_\theta)+D_1\nabla^2\right)W^{n}_1
	+\fhnEps W^{n}_2 = 0,				\nonumber\\
&&-\fhnEps^{-1}W^{n}_1 
 + \left(-\fhnEps\gamma+\omega(in+\@_\theta)
					\right)W^{n}_2 = 0, \quad n=0,1
						\label{EVPFHN}
\end{eqnarray}
for the unperturbed solution $U_j$, its angular velocity $\omega$
and the RFs $W^{n}_j$, $n=0,1$, $j=1,2$, $W^{0}_{1,2}\in\Real$, 
$W^{1}_{1,2}\in\Complex$ . System
\Eq{SWFHN}--\Eq{EVPFHN} should be supplied with normalisation and boundary
conditions, and discretized.  For discretisation we used rectangular
grids in Cartesian coordinates.

\subsection{Spiral wave problem}

Since the Response Functions are the solution of the adjoint 
linearized problem in the system of reference that is rotating with the 
angular velocity of the spiral itself, we need first to find the spiral 
wave solution in this system of reference, \ie\ we need to find the spiral 
wave solution {\it together} with its rotation angular velocity. So we have 
a nonlinear eigenvalue together with a boundary value problem.  

We solved this problem numerically on a square domain
$(x,y)\in\Square=[-L/2,L/2]\times[-L/2,L/2]$, for different $L$ from 25 to 50.
First, the spiral wave was initiated by solving a Cauchy problem for
\Eq{FHN} for initial conditions $u(x,y,0)=0.7\sign(x)$,
$v(x,y,0)=0.6\sign(y)$. When a stationary rotating spiral wave was
established, typically within time interval $t\in[0,T]$, $T\sim40$,
the resulting distribution $u(x,y,T)$, $v(x,y,T)$ was used as an
initial condition for the following system:
\begin{eqnarray}
\@_tu_1 &=& \fhnEps^{-1} \left( u_1 - u_1^3/3 - u_2\right)  + D_1\nabla^2 u_1
	+ \sum\limits_{j=x,y,\theta} C_j \@_j  u_1 , \nonumber\\
\@_tu_2 &=& \fhnEps \left(u_1+\beta-\gamma u_2\right)
	+ \sum\limits_{j=x,y,\theta} C_j \@_j  u_2 , \nonumber\\
\dot{C}_j &=& -q_jC_j 
	-  \frac{p_j}{A} \int\limits_{\Disk} (\@_tu_1\@_ju_1 + \@_tu_2\@_ju_2) \,\d{x}\d{y} ,
		\quad	j=x,y,\theta,		\label{FHNrot}
\end{eqnarray}
where $\@_\theta=x\@_y-y\@_x$, $\Disk=\{(x,y):\,x^2+y^2\le(L/2)^2\}$,
$A=\pi(L/2)^2$, $q_\theta=0$, and positive coefficients
$p_{x,y,\theta}$ and $q_{x,y}$ have been selected by experimentation.
Informally, the idea of this system, adopted with appropriate
modification from \cite{Biktashev-etal-1996}, is that the first two
equations are system \Eq{FHN} in a frame of reference moving with
speeds $-C_{x,y,\theta}$ in the $x$, $y$ and $\theta$ directions, and
the integrals in the evolution equations for $C_{x,y,\theta}$ are
``detectors of movement'' in those directions. So the movement of the
frame of reference is adjusted in such a way so as to make the
solution in this frame of reference is stationary. On the formal level,
it is straightforward to see that if solution of \Eq{FHNrot} converges
to a stationary state, then $u_{1,2}$ will satisfy \Eq{SWFHN} with
$\omega=-C_\theta$, neglecting the boundary conditions. 

We considered the problem \Eq{FHNrot} for the following set of
parameters: $D_1=1.0$, $\fhnEps=0.30$, $\beta=0.75$, $\gamma=0.50$,
which correspond to a rigidly rotating spiral wave solution
\cite{Winfree-1991}. Calculations were performed using the explicit
Euler method in time, central differences in space, with fixed time
step from $\stept = 3\cdot10^{-3}$ down to $\stept=5\cdot10^{-4}$ and
space step $\stepx = 0.5$, with Neuman boundary conditions on a
rectangular grid in Cartesian coordinates: $\@_xu_1(\pm
L/2,y)=\@_yu_1(x,\pm L/2)=0$. The frame of reference adjustment
parameters were chosen $q_{x,y}=1$, $p_{x,y}=7$ and $p_\theta=5$.

\myfigure{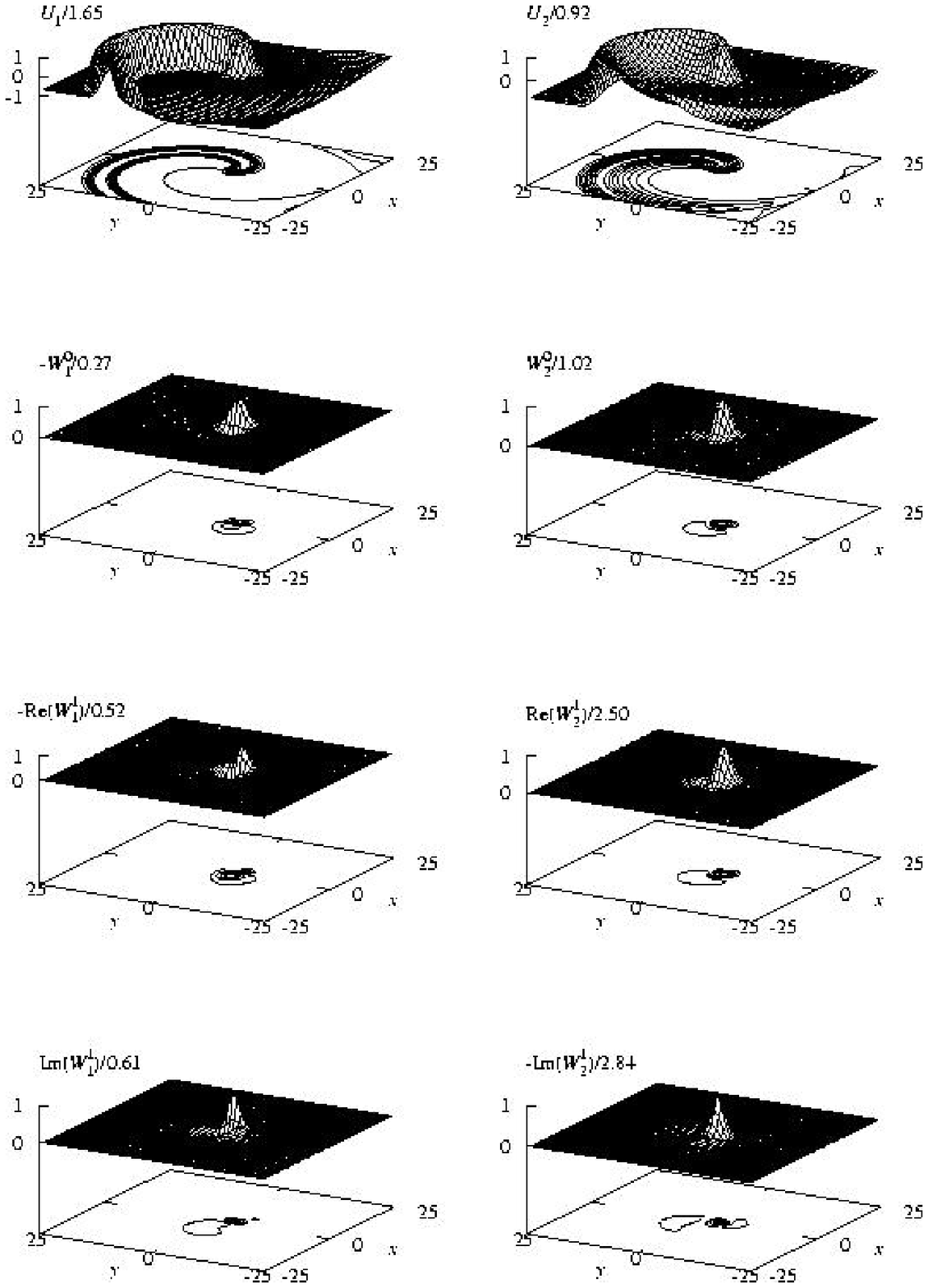}{
  Spiral wave solution and the response functions. 
  Parameters: $L=50$, $\stepx=0.5$, $\stept=5\cdot10^{-4}$, Neuman
  boundary conditions on $\@\Square$ for $\U$ and Dirichlet
  boundary conditions on $\@\Disk$ for $\W$. 
}{fhnrfs}

The result is a stable, stationary spiral (demonstrated on
\fig{fhnrfs}) in the system of reference, which rotates with the
angular velocity $\omega\approx0.32$ clockwise.  Having this angular
velocity and the spiral wave solution, it is possible to find the
corresponding response functions.

\subsection{The response functions problem}

The response functions $\W$ were calculated simultaneously with
finding the spiral wave solution for \Eq{FHNrot}, by solving the adjoint
linearized problems
\begin{eqnarray} 
\@_tw_1 &=& \fhnEps^{-1} (1-u_1^2) w_1 + \fhnEps w_2  + D \nabla^2 w_1
	- \sum\limits_{j=x,y,\theta} C_j \@_j  w_1 , \nonumber\\
\@_tw_2 &=&  -\fhnEps^{-1} w_1  - \fhnEps\gamma w_2
	- \sum\limits_{j=x,y,\theta} C_j \@_j  w_2 . \label{RFrot}
\end{eqnarray}
As $(u_1,u_2,C_1,C_2,C_3)$, solution of \Eq{FHNrot}, converges to
$(U_1,U_2,0,0,-\omega)$, solution of \Eq{EVPFHN}, then
$\w=(w_1,w_2)\T$, a typical solution of \Eq{RFrot}, is expected to
converge to
\begin{equation}
  \w(x,y,t) \approx 
    c_0 \RF_0(x,y)
    + \Re\left( c_1 \RF_1(x,y) \e^{-i\omega t} \right) ,
							\label{RFasymp}
\end{equation}
where $c_0\in\Real$ and
$c_1\in\Complex$, ideally, are constants depending on initial conditions,
but in real calculations can slowly change in time due to numerical
approximation. 
To obtain $\RF_{0,1}$, we calculate three solutions $\w^{(m)}$, $m=1,2,3$, 
to \Eq{RFrot} with three
linearly independent initial conditions. Assuming that, after a sufficiently
long time, these satisfy \Eq{RFasymp}, we can take their appropriate
linear combinations to satisfy \Eq{norm}. So for a given 
triplet $\w^{(m)}$, $m=1,2,3$, we are looking for constants
$P_{j,m}$, $j=\theta,x,y$, $m=1,2,3$, so that
\[ \W_j = \sum\limits_{m=1}^3 P_{j,m} \w^{(m)} \]
would satisfy biorthogonality condition \Eq{norm}
with respect to the set $\V_k=\@_k\U$, $k=\theta,x,y$. This requirement
implies that
\[ \sum\limits_{m=1}^3 P_{j,m} \< \w^{(m)}, \U_k \> = \delta_{j,k} , \]
\ie\ $\mx{P}=(P_{j,m})$ is an inverse to the matrix
$\mx{B}=(B_{m,k})$, where $B_{m,k}=\< \w^{(m)},\U_k \>$ are obtained
by integrating the given solutions of \Eq{RFrot} with the derivatives
of the solution of the nonlinear problem. Having thus found $\W_{x,y,\theta}$
after integrating \Eq{RFrot} for a time interval of certain $\tau$, we
used these $\W_{x,y,\theta}$ as three linearly independent initial conditions for
\Eq{RFrot} for a further time interval of length $\tau$.

Equation \Eq{RFrot} was integrated on the disk
$(x,y)\in\Disk\subset\Square$ with Dirichlet boundary conditions
$w_1(x,y)=w_2(x,y)=0$, $(x,y)\in\@\Disk$, which were implemented by
setting $w_{1,2}(x,y)=0$ for $(x,y)\in\Square\setminus\Disk$.  We
tried $\tau=1$ and $\tau=10$ with identical results; the solution
obtained by this procedure converged within time scale of $t\sim20$.
As $\V_k=\@_k\U$ are related to the Goldstone modes $\V_{0,\pm1}$ by
\Eq{Goldstone}, this gives relationship between functions $\W_{x,y,\theta}$
found in this way, with RFs, in the form $\W_0=-\W_\theta$ and $\W_1=-(\W_x-\i\W_y)$.

\Fig{fhnrfs} shows the components of the Response Functions 
obtained in this way. The most important
result is that all components are localized in a close vicinity of
the tip of the spiral. It can also be observed that the amplitude of
the second components of the RFs is higher than that of the first, so
controlling movement of the spiral wave via the second component, the
inhibitor, if it is pratically feasible, could be more efficient than
via the first component the activator. 

The accuracy of the solution has been checked by varying the
discretization steps $\stept$ and $\stepx$, the domain size $L$,
boundary conditions (Neuman vs Dirichlet, $\@\Disk$ vs $\@\Square$).
Based on such checks, we estimate that the accuracy of the solutions
is within a few percent. Further improvement of accuracy requires
decrease of $\stepx$ while keeping $L$ same or increasing, and due to
stability limitations of the fully explicit Euler time stepping, such
improvement is extremely costly if done within the same numeric
scheme, and requires a more advanced scheme, \eg\ fully
implicit/pseudospectral in the $\theta$ direction. 

\section{Conclusion}
The response functions are very important characteristics of the 
spiral wave, for they define the phenomenology of the spiral 
behaviour. Experiments and computer simulations, demonstrating the spirals insensitivity to distant events, implied that the vortices Response Functions must decay quickly with distance from the core. Such decay will guarantee the convergence of the 
integrals \Eq{forces} even for non-localized perturbations, for example 
caused by the simultaneous change of the properties of the whole 
medium. But the mathematical peculiarity of the idea presuming 
qualitatively different behaviour of eigenfunctions of the linear 
operator and its adjoint one resulted in a natural scepticism. 

Recently shown for the CGLE oscillatory medium localisation of the response functions in the vicinity of the vortices core left open the question on localisation properties of the spiral waves response functions in an excitable medium. The results obtained in this paper confirm the existence of effectively localized response functions of a spiral wave solution in 
the classical excitable FitzHugh-Nagumo model, at least for the particular set of the model parameters corresponding to a rigidly rotating spiral wave.  

\section{Acknowledgment}
Support by EPSRC grants GR/S43498/01 and GR/S75314/01 and the
Liverpool University Research Development Fund is acknowledged.

\bibliographystyle{agsm}
\bibliography{fhn} 
\end{document}